\documentclass{article}
\usepackage{spconf,amsmath,graphicx}
\usepackage{booktabs} 
\usepackage{subfigure}
\usepackage{url}
\usepackage[utf8]{inputenc}
\usepackage{makecell}
\usepackage{multirow}
\usepackage{textcomp}
\usepackage{graphicx}

\title{CONTEXTUAL BIASING FOR LLM-BASED ASR WITH HOTWORD RETRIEVAL AND REINFORCEMENT LEARNING}
\name{ YuXiang Kong$^{\dagger}$, JunFeng Hou$^{\dagger}$ \thanks{$^{\dagger}$Contributed equally.}, Jian Tang$^{\dagger}$, Bingqing Zhu, Jicheng Zhang, Shaofei Xue }
\address{
       Tongyi Lab, Alibaba Group, Hangzhou, China\\
       \{kongyuxiang.kyx, qianzhuo.hjf, yongran.tj, zhubingqing.zbq, zjc475228, shaofei.xsf\}@alibaba-inc.com
}

\begin{document}

\maketitle
\begin{abstract}
Large language model (LLM)-based automatic speech recognition (ASR) has recently achieved strong performance across diverse tasks, yet contextual biasing for named entities and hotwords under large vocabularies remains challenging. In this work, we propose a scalable two-stage framework that integrates hotword retrieval with LLM-ASR adaptation. First, we extend the Global–Local Contrastive Language–Audio pre-trained model (GLCLAP) to retrieve a compact top-$k$ set of hotword candidates from a large vocabulary via robustness-aware data augmentation and fuzzy matching. Second, we inject the retrieved candidates as textual prompts into an LLM-ASR model and fine-tune it with Generative Rejection-Based Policy Optimization (GRPO), using a task-driven reward that jointly optimizes hotword recognition and overall transcription accuracy.
Experiments on hotword-focused test sets show substantial keyword error rate (KER) reductions while maintaining sentence accuracy on general ASR benchmarks, demonstrating the effectiveness of the proposed framework for large-vocabulary contextual biasing.
\end{abstract}

    
\section{Introduction}\label{sec:intro}
End-to-end automatic speech recognition (ASR) systems have made substantial progress over the past decade.
More recently, the emergence of large language models (LLMs) has further advanced the field, enabling LLM-ASR frameworks \cite{bai2024seed,xu2025fireredasr,chu2024qwen2} that tightly integrate powerful language modeling with acoustic representations. 
By leveraging long-context reasoning and strong prior knowledge, LLM-ASR frameworks have achieved remarkable performance across a wide variety of tasks.

Despite these advances, contextual biasing remains a persistent challenge, especially for named entities and hotwords that are critical in many real-world applications \cite{yang2024promptasr,salemi2024lamp}.
Contextual biasing refers to enhancing recognition of task-relevant words given side information, such as contact lists and locations.
These items often carry high semantic or business value, yet appear infrequently in training data and may even be out-of-vocabulary. 

Traditional contextual biasing techniques such as WFST-based biasing graphs \cite{mohri2002weighted,gourav2021personalization}, shallow fusion with contextual language models \cite{hori2017advances,kannan2018analysis}, and class-based n-gram approaches \cite{wang2023incorporating} were primarily developed for non-LLM ASR architectures. 
While effective in conventional pipelines, they are not straightforward to integrate into LLM-ASR frameworks. 
More recent methods that supply contextual words through prompts or additional textual inputs to LLMs typically assume relatively small vocabularies \cite{lakomkin2024end,li2023prompting,flemotomos2025optimizing,hou2025ranking}.
Inspired by retrieval-augmented generation (RAG)~\cite{gao2023retrieval}, prior works~\cite{2025BR, 2024bias} adopt a two-stage paradigm: first retrieving candidate bias terms from a large-scale lexicon, and subsequently constructing a bias-aware textual prompt to condition the ASR model during decoding.
Moreover, they generally do not optimize directly for task-level metrics such as hotword recall or WER.
As a result, there is still a lack of a scalable framework that (i) efficiently selects relevant hotwords from massive vocabularies and (ii) systematically adapts LLM-ASR to use them effectively during decoding.

In this work, we address these limitations by decomposing contextual biasing for LLM-ASR into two tightly coupled subproblems: hotword retrieval and hotword-aware ASR adaptation. 
On the retrieval side, we extend the Global–Local Contrastive Language–Audio pre-trained model (GLCLAP) \cite{kong2025glclap} to enable efficient retrieval of hotword candidates from large vocabularies.
On the adaptation side, we adopt a reinforcement learning framework (RL) to fine-tune the LLM-ASR model.
We design a task-driven reward to jointly optimize hotword recognition and overall transcription fidelity. 
This RL-based optimization encourages the model to correctly recognize hotwords when present, avoid hallucinating them when absent, and maintain general transcription performance.
The results show that the proposed framework leads to significant WER reductions on hotword-centric evaluations, while preserving strong performance on general tasks.

The rest of this paper is organized as follows: Section \ref{sec:proposed} introduces our proposed system. Section \ref{sec:setups} shows our detailed setups. Section \ref{sec:experiments} presents the experimental results. Conclusions are drawn in Section \ref{sec:conclusion}.

\begin{figure*}[t]
  \centering
  \includegraphics[width=0.7\linewidth]{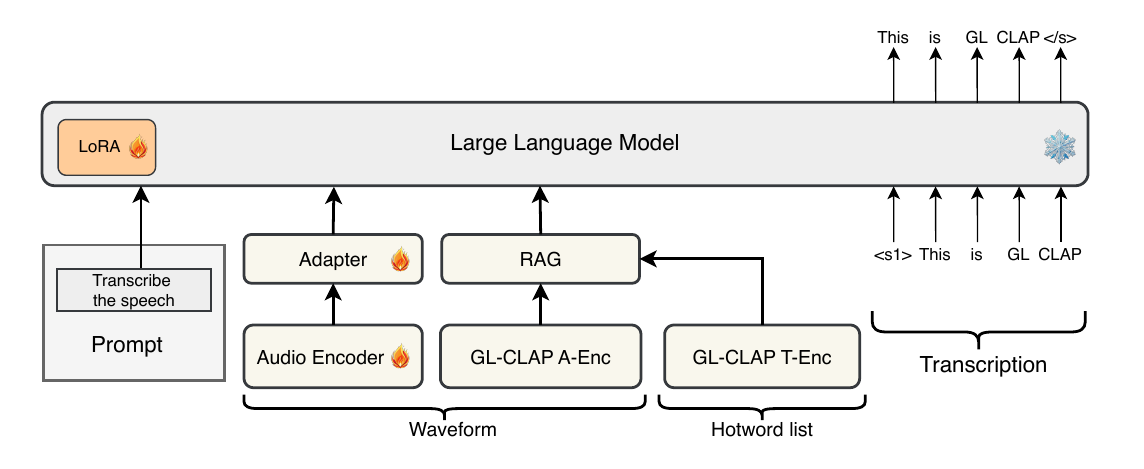}
  \caption{The proposed GLCLAP based contextual biasing LLM-ASR.}
  \label{fig:asr}
\end{figure*}

\section{Proposed System}\label{sec:proposed}
The framework of the proposed system is illustrated in Fig.~\ref{fig:asr},
consisting of the typical architecture of LLM-ASR and the
GLCLAP retrieval module.
\subsection{Framework Overview}\label{sec:framework}
\begin{figure}
  \centering
  \includegraphics[width=1.0\linewidth]{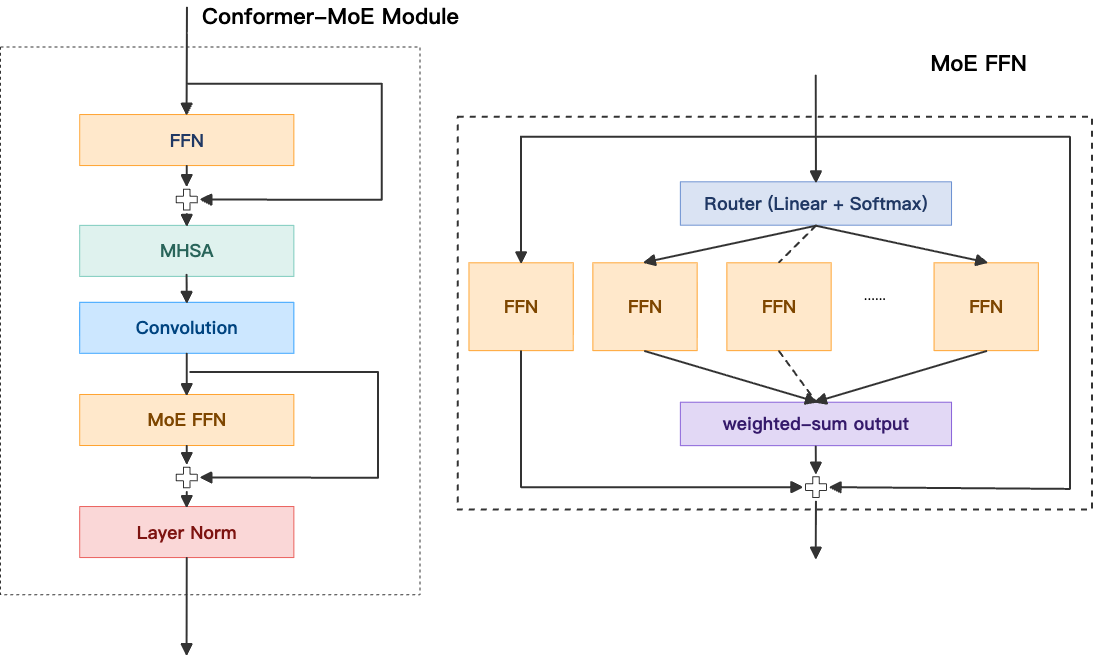}
  \caption{The structure of Conformer-MoE.}
  \label{fig:moe}
\end{figure}

First, we describe the main architecture of our model, an LLM-ASR network composed of an audio encoder, an adapter, and an LLM. In this work, we adopt Qwen2.5-7B as the LLM.
We extend the original Conformer encoder into a Conformer-MoE encoder, as illustrated in Fig.~\ref{fig:moe}, by replacing the second feed-forward network (FFN) module in each Conformer layer with a Mixture-of-Experts (MoE) structure.
We define $K_{C}$ candidate experts and employ a router to select $K_{S}$ of them for weighted aggregation, while one expert is dedicated as a shared expert.


\subsection{Enhanced GLCLAP for Hotword Retrieval}\label{sec:enhanced}
The GLCLAP retriever takes as input an audio signal $x$ and a set of candidate biasing words ${G=\{g_1,g_2,\dots,g_N\}}$, where $N$ denotes the size of the candidate list.
It consists of two components: an audio encoder (A-enc) and a text encoder (T-enc).
Given the input audio, A-enc extracts a fixed-dimensional audio embedding, denoted by ${h_{audio}}$.
Meanwhile, each biasing word ${g_i\in G}$ is encoded by T-enc into a semantic vector ${e_i}$, yielding a set of text embeddings ${E=\{e_1,e_2, \dots,e_N\}}$.
We then compute similarity scores between ${h_{audio}}$ and all embeddings in ${E}$, and select the top-${k}$ biasing words with the highest similarity scores to form a subset ${G'}$.
These selected biasing words are appended to the bias prompt to guide context-aware transcription.

Our system enhances overall context-aware recognition with three strategies: (i) integrating a data augmentation pipeline, (ii) incorporating a fuzzy matching strategy into GLCLAP training, and (iii) introducing a reinforcement learning phase into LLM-ASR training.

\subsubsection{Robustness-Aware Data Augmentation}
We construct a robustness-aware data augmentation (RADA) pipeline to reduce the size of the hotword vocabulary, since increasing the number of hotwords lowers recall and introduces more distractors.
The initial vocabulary is obtained by crawling domain-specific hotwords from the web.
For each biasing word ${g_i\in G}$, we synthesize speech using TTS (with text generated by an LLM when needed) and decode it with the ASR system to check whether the original LLM-ASR can already recognize this hotword. 
If it is reliably recognized, the word is removed from the hotword vocabulary; otherwise, it is retained.

\subsubsection{Fuzzy Matching Strategy}
Real-world applications often cannot rely on exact matches to the hotword vocabulary; otherwise, it would become prohibitively large. However, during GLCLAP training, hotwords are enforced via strict lexical matching, which is inconsistent with deployment scenarios where users may produce inflected forms, paraphrases, or partial mentions of target terms.

To better match practical usage and improve end-to-end performance, especially for partial or fuzzy matches, we augment the training data with embeddings of generated contextual sentences and deliberately perturbed variants of biasing words (e.g., “Tongyi abc” instead of “Tongyi”). This fuzzy matching strategy captures real-world ambiguity and improves robustness.

\subsubsection{Contextual ASR through RL-Guided Discrimination}
To suppress false positives, we leverage the ASR model’s ability to discriminate biasing words during decoding. The model is trained with structured prompts of the form:
“⟨Audio⟩ Transcribe the audio into text. These biasing words you may use: ⟨$g_1$⟩ ⟨$g_2$⟩ ... ⟨$g_k$⟩”,
where the provided biasing word list deliberately includes irrelevant or distractor terms to discourage overreliance on the biasing words.

In addition to the data-level operations mentioned above, we introduce Generative Rejection-Based Policy Optimization (GRPO)~\cite{shao2024deepseekmath}, a reinforcement learning method applied during LLM-ASR training to further enhance the model’s discrimination capability. Our reward function is designed to jointly optimize multiple objectives, including:
\begin{itemize}
    \item a match reward: if a candidate biasing word appears in both the model output and the reference label, or in neither of them, the reward is 1; otherwise, the reward is 0; 
    \item a WER-based reward: $\text{reward} = 1 - \text{WER}$, ensuring transcription accuracy.
\end{itemize}

To further improve performance, we also adopt a joint beam search strategy during inference~\cite{bai2024seed}. By jointly decoding context-free and context-conditioned hypotheses, we preserve the effectiveness of retrieval-augmented generation (RAG) while reducing hallucinations caused by irrelevant biasing words.




\section{Setups}\label{sec:setups}
\subsection{Dataset and Metrics}

\subsubsection{Context-Biasing Training Datasets}
Our LLM-ASR system is trained on several million hours of speech data in total. Here we focus on the context-biasing data. After base LLM-ASR training, we further fine-tune the model on 2M utterances associated with hotwords and/or contextual history, mainly generated by the RADA process. Utterances with and without hotwords/context are mixed with a ratio of 8:1 in favor of non-biased data.
For utterances containing hotwords, each utterance has 1–10 hotwords. About half of them contain the correct target hotwords and the other half do not, yielding a 1:1 ratio between positive and negative hotword instances.

\subsubsection{GLCLAP Fine-Tuning Data}
In the first stage, we follow the training configuration in the original GLCLAP paper. The model is trained on large-scale general ASR data, where text labels are obtained by randomly cropping short phrases from full transcripts.
For the second-stage fine-tuning of the GLCLAP retriever, we construct a domain-targeted dataset of about 250k audio–text pairs to better align retrieval with our biasing-word domains.

\subsubsection{Hotword Vocabulary}
The hotword vocabulary is compiled from web data, focusing on two domains: medical and media (film and television). The initial list is then filtered using the RADA strategy. The vocabulary size is reduced from about 600k entries to 98k hotwords after RADA-based filtering.

\subsubsection{Evaluation}
We build two targeted test sets, \textbf{Media} and \textbf{Medical}, each containing 240 utterances mainly collected from bad cases. Each utterance is manually annotated with ground-truth biasing words, which are added to the biasing-word list.
We also construct a regression test set, \textbf{General Task}, consisting of about 5k standard ASR utterances.

We use different metrics at different stages.
For GLCLAP, we use recall as the primary metric. Unlike the conventional definition, a hotword is considered successfully recalled if the annotated hotword is a substring of the detected hotword, rather than an exact match. This accounts for lexical variants in realistic hotword lists (e.g., “qwen2”, “qwen2.5”, “qwen3” for the underlying hotword “qwen”).
For the final ASR evaluation, we use (i) sentence-level recognition accuracy (SACC), and (ii) keyword error rate (KER). For KER, if any annotated keyword of an utterance is missing in the ASR output, that keyword is counted as an error.

\subsection{Model Configurations}
Our LLM-ASR model has 10.5B parameters in total, with 8.7B active parameters during inference. The architecture consists of a 3.5B-parameter encoder and an LLM decoder. The encoder has a CNN front end followed by 20 Conformer-MoE layers. The CNN performs 4× temporal downsampling, and the resulting features are fed into the Conformer-MoE stack. Each MoE layer uses a 3-of-8 expert routing scheme with a hidden size of 3584.
The encoder output is further processed by 2× frame-level downsampling with concatenation, then passed through a two-layer linear adapter before being fed into the LLM. During base training, only the LLM is updated using LoRA~\cite{hu2022lora}, with rank 64 and alpha 32.
During context-biasing training, we use a learning rate of 1e-5 and jointly update the audio encoder, adapter, and LLM LoRA parameters.
During GRPO training, we use the same learning rate (1e-5) but freeze the encoder and adapter, updating only the LLM LoRA parameters. We set the KL-divergence weight to 0.04 and generate six responses per step. 

\section{Experiments}\label{sec:experiments}
\subsection{Retrieval Performance Evaluation}

We first evaluate GLCLAP-based hotword retrieval on the Media and Medical test sets. As shown in Table~\ref{tab:mapping1} and Table~\ref{tab:mapping2}, both Robustness-Aware Data Augmentation (RADA) and fuzzy matching contribute positively to overall performance. Specifically, the original hotword vocabulary contains approximately 600k entries, which is reduced to 98k entries after applying RADA.
Fuzzy matching not only aligns more closely with our evaluation metric, but also better reflects real-world application scenarios. In addition, the results show that the recall rate consistently increases as the top-$k$ value grows.

\begin{table}[th]
\caption{Recall rate of Extended GLCLAP methods on Medical testset for various top-$k$.}
\centering
  \begin{tabular}{ l c c c c }
    \toprule
\multicolumn{1}{c}{\textbf{Recall(\%)}} & \textbf{top-1} & \textbf{top-2} & \textbf{top-5} &\textbf{top-10}  \\
    \midrule
    GLCLAP & 23.67 & 29.77 & 39.31 & 42.37 \\
    +RADA & 45.80 & 53.44 & 62.98 & 67.56 \\
    +Fuzzy Matching & \textbf{64.89} & \textbf{74.43} & \textbf{80.54} & \textbf{85.12} \\
    \bottomrule
\end{tabular}
\label{tab:mapping1}
\end{table}

\begin{table}
\caption{Recall rate of Enhanced GLCLAP methods on Media testset for various top-$k$.}
\centering
  \begin{tabular}{ l c c c c }
    \toprule
\multicolumn{1}{c}{\textbf{Recall(\%)}} & \textbf{top-1} & \textbf{top-2} & \textbf{top-5} &\textbf{top-10}  \\
    \midrule
    GLCLAP & 27.98 & 35.78 & 40.37 & 43.58 \\
    +RADA & 59.63 & 66.51 & 72.94 & 76.61 \\
    +Fuzzy Matching & \textbf{83.03} & \textbf{89.00} & \textbf{92.20} & \textbf{93.58} \\
    \bottomrule
\end{tabular}
\label{tab:mapping2}
\end{table}

\subsection{ASR Performance of Enhanced GLCLAP}
The previous experiment examined how the recall rate varies as a function of the top‑$k$ value. In Table~\ref{tab:topk}, we combine GLCLAP with the LLM‑ASR model and report the final recognition performance in terms of SACC and KER. The Base column shows the results of the context‑aware fine‑tuned LLM‑ASR model decoded without any bias prompt.
We observe that although the recall rate continues to increase as top‑$k$ grows, KER and SACC on the hotword test sets do not improve monotonically. This is because a larger top‑$k$ introduces more distractor candidates to the LLM‑ASR model, thereby increasing interference.
On the General Task, most results are slightly worse than the baseline without hotwords. Considering both hotword test sets together, we conclude that top‑$2$ is a more appropriate choice.



\begin{table}[th]
\caption{ASR comparison of Enhanced GLCLAP on different top-$k$.}
\centering
\resizebox{1.1\columnwidth}{!}{
  \begin{tabular}{ l l | c c c c c}
    \toprule
\multicolumn{1}{c}{\textbf{ }}  
                                      &   &  \textbf{base}  & \textbf{top-1}
                                         & \textbf{top-2}
                                         & \textbf{top-5} & \textbf{top-10}\\
    \midrule
\multirow{3}{*}{\textbf{Media}}  & -recall(↓)  & -- & 83.03 & 89 & 92.2 & 93.58\\
&-KER(↓) & 22.63 & 12.91 & 11.99 & 13.37 & 15.21 \\
&-SACC(↑)  & 72.99 & \textbf{81.04} & 80.57 & 79.15 & 77.25 \\
\midrule
\multirow{3}{*}{\textbf{Medical}}  &-recall(↓) & -- & 64.89 & 74.43 & 80.54 & 85.12  \\
&-KER(↓) & 23.31 & 17.83 & 15.89 & 15.51 & 15.89\\
&-SACC(↑)  & 71.56 & 75.83 & 77.73 & \textbf{78.2} & 77.73 \\
\midrule
\textbf{General}  &  -SACC(↑)  & 77.49 & \textbf{77.52} & 77.36 & 77.42 & 77.39\\
    \bottomrule
\end{tabular}
}
\label{tab:topk}
\end{table}


\begin{table}[th]
\caption{Performance of applying GRPO training.} 
\centering
  \begin{tabular}{ l l | c c c }
    \toprule
\multicolumn{1}{c}{\textbf{ }}  
                                       &  &  \textbf{base}  &
                                         \textbf{+EGLCLAP} &
                                         \makecell{\textbf{+EGLCLAP} \\ \textbf{+GRPO}} \\
    \midrule
\multirow{2}{*}{\textbf{Media}}  & -KER(↓) & 22.63 & 11.99 &  \textbf{8.29} \\
&-SACC(↑) & 72.99 & 80.57 &  \textbf{85.78} \\
\midrule
\multirow{2}{*}{\textbf{Medical}}  & -KER(↓) & 23.31 & 15.89 &  \textbf{11.24} \\
&-SACC(↑)  & 71.56 & 77.73 &  \textbf{84.83} \\
\midrule
\textbf{\textbf{General}}  & -SACC(↑)  & 77.49 & 77.36 &  \textbf{78.16} \\
    \bottomrule
\end{tabular}
\label{tab:grpo_v}
\end{table}

\subsection{ASR Performance of Applying GRPO Training}
Table~\ref{tab:grpo_v} compares the results of applying GRPO training after context‑aware fine‑tuning. It shows that, on top of the +top‑2 configuration, introducing GRPO yields clear performance gains in KER on both the Media and Medical sets. Moreover, thanks to the accuracy-based reward used in GRPO, the sentence accuracy on the General Task is also significantly improved.



\section{Conclusions}\label{sec:conclusion}
In this paper, we proposed a scalable two-stage framework for contextual biasing in LLM-ASR with large hotword vocabularies. An enhanced GLCLAP retriever, equipped with robustness-aware data augmentation and fuzzy matching, selects a compact top‑$k$ hotword set per utterance. The retrieved hotwords are then injected as structured prompts into an LLM-ASR model, which is further optimized with GRPO using a combined bias-word match and WER-based reward.

Experiments on Media and Medical hotword test sets show consistent recall gains and substantial KER reductions, while sentence accuracy on a general ASR task is preserved. These results demonstrate that combining retrieval-based hotword selection with RL-guided LLM-ASR adaptation is effective for large-vocabulary contextual biasing.
Future work will explore extending this framework to multilingual settings and tighter joint optimization of retrieval and decoding.



\bibliographystyle{IEEEbib}
\bibliography{strings,refs}

\end{document}